\documentclass[conference,a4paper]{IEEEtran}
\usepackage{amsfonts}
\usepackage{amssymb}
\usepackage{amsmath}
\usepackage[printonlyused,nolist]{acronym} 
\usepackage{booktabs}
\usepackage{tabularx}
\usepackage{array,ragged2e,calc}
\usepackage{graphicx}
\usepackage{lipsum}
\usepackage{multirow}
\usepackage{xcolor}
\usepackage{url}
\usepackage{float}
\usepackage{breakurl}

\usepackage{algorithm}
\usepackage{algpseudocode}
\usepackage[font=footnotesize]{caption}
\IEEEoverridecommandlockouts

\title{Quality of Service Aware Traffic Management for Aircraft Communications}

\author{\IEEEauthorblockN{David~Tomi\'c\IEEEauthorrefmark{1},  Sandra~Hofmann\IEEEauthorrefmark{2},  Mustafa~Ozger\IEEEauthorrefmark{1}, Dominic~Schupke\IEEEauthorrefmark{2}, and Cicek~Cavdar\IEEEauthorrefmark{1}}
	\IEEEauthorblockA{\IEEEauthorrefmark{1}Division of Communication Systems, KTH Royal Institute of Technology, Stockholm, Sweden \\Email: \{dtomic, ozger, cavdar\}@kth.se}
	\IEEEauthorblockA{\IEEEauthorrefmark{2} Airbus, Central Research and Technology, Munich, Germany
		\\Email: \{sandra.s.hofmann, dominic.schupke\}@airbus.com}
	
}

\begin{document}
\maketitle
\begin{abstract}
\boldmath
In-flight Internet connectivity is a necessity for aircraft passengers as well as aircraft systems. It is challenging to satisfy required quality of service (QoS) levels for flows within aircraft due to the large number of users and the highly varying air to ground (A2G) link capacities composed of satellite and direct air to ground communication (DA2GC). To represent service quality variations, we propose models for the generated traffic flows from aircraft and variations in A2G links. We present three different forwarding schemes based on priority, delay requirements and history of the dropped flows metrics. Forwarding schemes schedule the flows in real time by choosing either satellite or direct air to ground link depending on the delay and capacity requirements of flows to maximize the number of accepted flows with required QoS guarantees in terms of dropped packets and delay. Also, the effect of local caching is studied to fully satisfy the QoS requirement of flows in simulated flights. We implement the forwarding procedures and caching in ns-3 and test their performance in a current connectivity scenario of 100 Mbps capacity for both the satellite spot and ground base station in a one-hour flight. Our study shows that although the forwarding procedure based on a combination of priority and delay requirement has relatively better performance than the other schemes, which are based on priority only and weighted average of all metrics, in dropped packet percentage and delay, the current connectivity setup is not able to satisfy all QoS requirements. Furthermore, at least 0.9 cache hit rate is required to satisfy all flows for at least 50\% of simulated flights.

\end{abstract}
\setlength{\textfloatsep}{4pt}

\begin{IEEEkeywords}
aircraft communications, traffic management, quality of service, ns-3
\end{IEEEkeywords}

\section{Introduction}\label{sec:intro}

The ambitious goal of providing broadband Internet connectivity anywhere and anytime is also valid for commercial flights and passengers on-board \cite{NGMNAlliance2015}. \ac{IFBC} can be provided via \ac{DA2GC}, \ac{SA2GC} and \ac{A2AC} \cite{network_michal}.  

Generated flows in a passenger aircraft have different characteristics in terms of traffic shape and \ac{QoS} requirements. Extending \cite{arinc811}, each traffic flow in an aircraft network can be originated from the following six application domains:  \ac{ACD} - safety essential systems managing the aircraft and the flight, \ac{AISD} - safety non-essential systems, destined for maintenance and crew communication, \ac{PODD} - passengers devices such as smartphones and laptops, \ac{MTC} - sensors generating information to be sent to the ground, \ac{FRD} - devices capturing flight data and cockpit voice, \ac{HMD} - devices capturing thousands of aircraft health parameters for predictive maintenance.



Requirements of the listed domains on-board can be varying, and each traffic flow can be forwarded over either of the following communication links: The first one, \ac{DA2GC}, depends on a ground network of \acp{BS} with varying cell sizes and activity levels. The aircraft is directly connected to a \ac{BS} on the ground. As it moves, it is handed over from one \ac{BS} to another, receiving a share of the total \ac{BS} capacity. The second option is \ac{SA2GC}. In this paper, we focus on geostationary satellites providing access to the ground core network, however, low Earth orbit satellites can be considered for future scenarios. Handover and capacity sharing apply here too, however, satellite spots are much larger. Hence, less handovers and a smaller share of capacity can be expected. Additionally, \ac{SA2GC} causes a higher packet delay than \ac{DA2GC}. 

Link conditions for \ac{SA2GC} and \ac{DA2GC} are changing due to aircraft movement, and handovers to ground \ac{BS} cells and satellite spots with varying activity levels. They are also affected by the sharing of transmission capacity offered in a cell or spot with other aircraft. Hence, with traffic flows coming from/going to the devices within the aircraft, it is important to find which one should be forwarded over \ac{SA2GC}, and which one over \ac{DA2GC}. Given the link and traffic conditions over the flight time, a forwarding scheme decides on the allocation of in-flight traffic flows to the available links. Queuing policies to decide which flows need to be prioritized also affect the performance of the traffic management. However, our focus is based on the sole effect of forwarding schemes to satisfy \ac{QoS} requirements of the flows for the traffic management. Furthermore, it needs to decide on flows that need to be dropped in an overload situation. How to forward these flows according to their \ac{QoS} requirements is an important research question we aim to address. 


Apart from the connectivity options, caching is a promising complement to satisfy the \ac{QoS} levels of flows in aircraft. It can enhance the \ac{QoS} under limited \ac{SA2GC} and \ac{DA2GC} link performance. Hence, local caching will be a crucial requirement in future \ac{IFBC} networks \cite{Lufthansa}. A cache can be filled with content which is frequently fetched from the Internet by users, or with content placed into the cache beforehand. Both options require a strategy on selecting the right content. 
If the filled content is chosen based on a strategy to predict user demand, and the cache size is big enough, high cache hit rates could be expected. The cache hit rate describes the percentage of requests for the content which are satisfied by the cache.

In this paper, we propose traffic models to simulate behavior of the flows generated within the aircraft. We also model the varying link capacities for \ac{DA2GC} and \ac{SA2GC}. Given models for the flows and varying link capacities, we create a simulation framework using \ac{ns-3} and OpenFlow. Along this framework, we propose a forwarding logic for \ac{QoS} aware traffic management. The aim of this paper is to study the effects of forwarding schemes to maximize the number of accepted flows with required \ac{QoS} guarantees. Our forwarding schemes depend on the weighted sum of priority, delay requirement and traffic flow history (number of times a flow was dropped). We test their \ac{QoS} performance in terms of the percentage of dropped packets for passenger flows of different classes and the percentage of total simulation time spent on \ac{SA2GC} link by VoIP flow. We also study the percentage of flights in which QoS requirements of all flows are satisfied over a number of simulations with caching.

The remainder of the paper is organized as follows. Section \ref{sec:model} provides our system model explaining in-flight traffic model and \ac{SA2GC} and \ac{DA2GC} link model. Section \ref{sec:forwarding algorithm} describes the forwarding algorithm. Section \ref{sec:sim_setup} explains the simulation setup. Section \ref{sec:results} provides the results of the simulation study. Finally, Section \ref{sec:conclusion} concludes our paper. 

\section{System Model}\label{sec:model}


\subsection{Traffic Model}\label{sec:trafficmodel}

Traffic originating from the aircraft network is already presented in today's commercial aircraft \cite{traffic_model}, which are \ac{ACD}, \ac{AISD}, \ac{HMD} and \ac{FRD}. Modeling \ac{MTC} traffic for an aircraft is still an open research question. Overall however, these flows are assumed to be represented by one device respectively, which is generating the expected traffic volumes. 

\begin{table}
	\caption{Application domains with the respective priority and delay requirement} 
	\centering
	\scriptsize
	\begin{tabularx}{0.45\textwidth}{c|c|c}\hline
		\textbf{Application Domain} & \textbf{Priority } & \textbf{Delay} \\
		\hline \hline
		\textbf{\ac{ACD}}   &5 &3 \\ \hline
		\textbf{\ac{AISD}}   &5 &3 \\ \hline\textbf{\ac{HMD}}   &5 &3 \\ \hline\textbf{\ac{FRD}}   &5 &3 \\ \hline \textbf{PODD - VoIP, Video, Web - First }  &4 & \{3, 2, 1\} \\ \hline
		\textbf{PODD - VoIP, Video, Web - Business  }  &3 & \{3, 2, 1\} \\ \hline
		\textbf{PODD -  VoIP, Video, Web - Economy  }  &2 & \{3, 2, 1\} \\ \hline
		\textbf{\ac{MTC} }   &1 &1 \\ \hline
	\end{tabularx}
	\label{tab:priority_delay}
\end{table} 

For the \ac{PODD} flows, types of traffic and volumes depend on the applications used by passengers, the percentage of applications used by passengers, and the total number of active passengers actively using an application; henceforth user activity. \ac{VoIP}, Video and Web applications are selected to be representative for the passenger traffic. Consequently, the usage ratio of \ac{PODD} applications is assumed to be $20$\% for VoIP and Web, $60$\% for Video. We assume three travel classes: first; business; and economy. They are considered to model different types of \ac{QoS} levels. 



The generated traffic flows have different levels of priorities and delay requirements, which are presented in Table \ref{tab:priority_delay}. Note that MTC is assigned the lowest priority, due to the assumption that data from this domain can be saved and analyzed after flights. Priority is defined to scale from $1$ to $4$ (lowest to highest), and delay requirement from $1$ to $3$ (lowest to highest).



According to \cite{ETSI}, \ac{VoIP} flows follow an ON/OFF traffic generation pattern, with exponentially distributed ON and OFF phases. The traffic generated in the ON phases has a \ac{CBR}. The same applies for Web flows with different characteristics. Video flows  are shown to have a continuous traffic generation, with a varying bit rate depending on the content represented in each video frame \cite{cisco3}. Data rates taken for VoIP, Video and Web are based on values given in \cite{cisco, cisco2, ns-3}. 
The \ac{IFBC} network is based  on packet switching.  Traffic is generated according to Table \ref{tab:traffic_analysis} showing the traffic behavior for each flow. 

\begin{table}
	\caption{Flow behavior and data rate for each application.} 
	\centering
\scriptsize
	\begin{tabularx}{0.49\textwidth}{X|X|X|X|X}\hline
		
		\textbf{Application} & \textbf{Traffic  pattern} & \textbf{Mean duration, distribution/ ON phase}& \textbf{Mean duration, distribution/ OFF phase}&\textbf{Data rate, type}\\
		\hline \hline 
		\textbf{VoIP (\ac{PODD})}   & ON/OFF&$3$ seconds,\ \ \ \ \ \ \ \ \ \ \   exponential&$3$ seconds, \ \ \ \ \ \ \ \ \ \ \  exponential & $15$ kbps, CBR\\ \hline
		\textbf{Video (\ac{PODD})}   & Continuous& - &- & $4$ Mbps, varying\\ \hline
		\textbf{Web (\ac{PODD})}   & ON/OFF&$5$ seconds, \ \ \ \ \ \ \ \ \ \ \   exponential&$30$ seconds,\ \ \ \ \ \ \ \ \ \ \   exponential & $3$ Mbps, CBR\\ \hline
		\textbf{ACD}   & Continuous &-& - & $0.08$ kbps, CBR\\ \hline
		\textbf{AISD}   & Continuous& - & - & $100$ kbps, CBR\\ \hline
		\textbf{HMD}   & Continuous& - & - & $0.6$ kbps, CBR\\ \hline
		\textbf{FRD}   & Continuous& - & - & $136$ kbps, CBR\\ \hline
		\textbf{MTC}   & Continuous& - & - & $3$ Mbps, CBR\\ \hline
	\end{tabularx}
	\label{tab:traffic_analysis}
\end{table}

%

\subsection{Link Model}\label{sec:linkmodel}
\begin{figure}[!t]
	\centering
	\includegraphics[width=.47\textwidth]{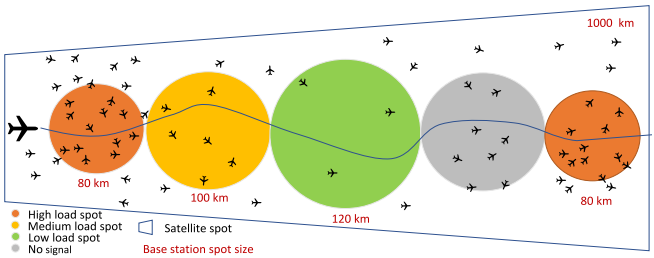}
	\caption{Aircraft traversing spots with different activity levels.}\label{fig:linkmodel}
\end{figure}

The available capacity per aircraft at any given time depends on the link conditions. During a flight, an aircraft is passing through \ac{DA2GC} cells and \ac{SA2GC} spot beams. For simplicity, we use the term spot for both cases. The number of aircraft within one spot changes over time. We define the activity level over a spot as having low load, medium load and high load, depending on the number of aircraft within the spot. All aircraft in one spot need to share the total available capacity, hence, less capacity per aircraft can be achieved in a high load spot. Over one flight, an aircraft is traversing multiple spots as illustrated in Figure \ref{fig:linkmodel}.

To obtain the number of aircraft within any type of spot, we analyzed real aircraft flight data over one week from \cite{FlightRadar} using MATLAB. We conduct this analysis for flights over Europe. From \cite{EAN}, we model a spot radius ranging between $80$ km and $150$ km in steps of $5$ km for \ac{DA2GC}. For \ac{SA2GC}, we assume $1000$ km for Europe, as actual spot sizes differ substantially depending on the geographic position and the satellite operator. To obtain aircraft densities, we assumed rectangular spots without loss of generality. As the satellite spot is much bigger, we assume multiple satellite operators, covering the same area to allow a fair comparison of achievable throughput. Otherwise, the achievable throughput for \ac{SA2GC} would be insignificant. In our case, we assume $10$ satellite operators.

We analyze the number of aircraft in a given spot every $15$ minutes for each spot size. The analysis shows that there are big differences between maximum and minimum numbers of aircraft, i.e., during the night most spots are at low load. From this analysis, we model $3$ different activity levels. For each spot size, a representative number of aircraft for each activity level is calculated. Exemplary values for some spot sizes can be found in Table \ref{tab:activity}. We define activity of spots as follows: Low load spot - number of airplanes lies at the midpoint between $5^{\mathrm{th}}$ percentile and median; Medium load spot - number of airplanes lies at the midpoint between median and $95^{\mathrm{th}}$ percentile; High load spot - number of airplanes lies at the midpoint between $95^{\mathrm{th}}$ percentile and maximum.

Additionally, we analyzed the number of leaving and arriving aircraft per spot every $3$ minutes. We conclude that for all spot sizes, the $5^{\mathrm{th}}$ and $95^{\mathrm{th}}$ percentiles range between $-4$ and $4$. Hence, we model up to $4$ aircraft coming or leaving every three minutes. The time spent in each spot is determined by its size and the aircraft speed – which is assumed to be an average of $900$ km/h.

\begin{table}
\caption{Number of aircraft in different spot types for selected spot sizes.} 
\centering
	\scriptsize
\begin{tabularx}{0.4\textwidth}{l||l|l|l}\hline
\textbf{Spot Size [km]} & \textbf{Low Load}&\textbf{Medium Load}&\textbf{High Load}\\
\hline \hline
\textbf{80}   & 1&10&55\\ \hline
\textbf{100}  & 2&11&61\\ \hline
\textbf{120}  & 2&14&80\\ \hline
\textbf{150}  & 3&16&103\\ \hline
\textbf{1000} & 4&28&67\\ \hline
\end{tabularx}
\label{tab:activity}
\end{table} 

The experienced spot sizes and activity levels of a flight depend on the airplane route. Since routes and times differ for each flight, for our model we assume that a random set of spots is passed during a flight. Within this flight also random spot sizes and activity levels are observed. This approach allows a general analysis independent of a specific route or deployment. These numbers can, however be adjusted if a specific operator or route should be analyzed.  

\subsection{Network Architecture}
\begin{figure}[!h]
	\centering
	\includegraphics[width=.49\textwidth]{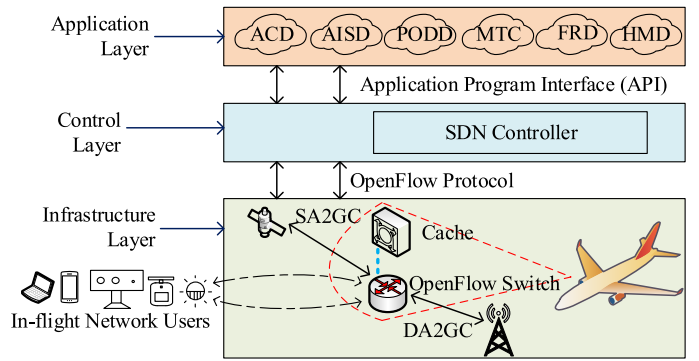}
	\caption{SDN architecture with in-flight network users.}\label{fig:net_arch}
\end{figure}

Figure \ref{fig:net_arch} shows the \ac{SDN} architecture with in-flight network users, which are connected to an OpenFlow switch. Applications inside the aircraft in the application layer (originated from the mentioned six domains) demand resources according to their \ac{QoS} requirements from SDN controller using application programming interfaces (API). Our OpenFlow switch resides in the infrastructure layer. \ac{DA2GC} and \ac{SA2GC} are two options for the OpenFlow switch to decide on which option to forward the flows. If a local cache is deployed on the aircraft, the required content can be fetched depending on the cache hit rate. Web and Video content can be cached to satisfy service requirements in case of capacity shortage.

The forwarding logic is implemented at the SDN controller in the control layer as seen in Figure \ref{fig:net_arch}. The developer of a new forwarding algorithm can implement it as an application, which is making use of those control plane/layer instructions. The instructions modify the forwarding tables in the switch, to effectively forward traffic by the OpenFlow switch in the infrastructure layer. 



\section{Forwarding Algorithm}
\label{sec:forwarding algorithm}

Main methods for the forwarding algorithm are provided in Algorithm \ref{alg:pseudo}. Inputs to the algorithm are a set of traffic flows with different priority, rate and delay requirements, change in the available capacity of \ac{DA2GC} and \ac{SA2GC} links, and change in the traffic flow data rate. According to these inputs, the following actions are performed: flows are  forwarded over DA2GC or SA2GC, offloaded from DA2GC to SA2GC (and vice versa) or in the worst case dropped. 

The decision variable to take the stated actions is the forwarding scheme value (FSV) for the flows, which are calculated as follows for different schemes: 
\begin{itemize}
	\item Forwarding scheme 1: Priority,
	\item Forwarding scheme 2: $0.5 \times$Priority + $0.5 \times$ delay requirement,
	\item Forwarding scheme 3: $0.5 \times$ Priority + $0.25 \times$delay requirement + $0.25 \times$number of times the flow was dropped.	
\end{itemize} Forwarding scheme 1 can be regarded as the baseline to measure how other terms affect the performance of forwarding decisions. 
The procedures used in Algorithm \ref{alg:pseudo} are given in Algorithm \ref{alg:procedures}. For example, if we have an incoming flow and the data rate requirement is satisfied with the available capacity of DA2GC, our algorithm searches for another flow with lower FSV to offload to SA2GC or even drop it. The same logic is followed in the other procedures, in which FSV is the main parameter to take decisions. For the forwarding algorithm, directions of the flows are considered for both uplink and downlink. Hence, the decisions are performed based on bidirectional flows for the development of this concept. 

 



\begin{algorithm}[t]

\scriptsize
	\caption{Pseudocode for Forwarding Algorithm.}
	\begin{algorithmic}[1]
		\While {True}
		\If {$\exists$ (incoming traffic flow)}
		\State {\textit{forward over DA2GC (incoming flow F)}}
		\ElsIf {$\exists$ change in DA2GC capacity}
		\State {\textit{change in capacity over DA2GC(new capacity)}}
		\ElsIf {$\exists$ change in SA2GC capacity}
		\State {\textit{change in capacity over SA2GC (new capacity)}}
		\ElsIf {$\exists$ change in traffic flow data rate}
		\If {flow is forwarded over DA2GC}
		\State {\textit{change in capacity over DA2GC (new capacity)}}
		\Else 
		\State {\textit{change in capacity over SA2GC (new capacity)}}
		\EndIf
		\EndIf
		\EndWhile
	\end{algorithmic}
	\label{alg:pseudo}
\end{algorithm}

\begin{algorithm}[t!]

	\scriptsize
		\caption{ Procedures in Algorithm \ref{alg:pseudo}.}
	\begin{algorithmic}[1]		
		
		\Procedure {Forward Over DA2GC}{incoming flow F}
		\If {Data Rate (DR) of F $\leq$ DA2GC Available Capacity (AC)}
		\State {Forward F over DA2GC}
		\Else
		\While {$\exists$ a flow with lower forwarding scheme value (FSV) on DA2GC link}
		\State {Offload that flow to SA2GC } 
		\If {DR of F $\leq$ DA2GC AC}
		\State {Forward F over DA2GC}
		\EndIf
		\EndWhile
		
		\EndIf
		\EndProcedure
		
		\Procedure {change in capacity over DA2GC}{new capacity}
		\If {new capacity $>$ old capacity}
		\While {DR of the flow with the highest FSV on SA2GC $\leq$ remaining capacity over DA2GC}
		\State {forward the flow with the highest FSV from SA2GC over DA2GC}
		\EndWhile
		\Else
		\While {remaining capacity (RC) over DA2GC $\leq$ 0}
		\While {RC over SA2GC $\leq$
			DR of the flow with the lowest FSV flow on DA2GC}
		\State {drop the flow with the lowest FSV flow from SA2GC}
		\EndWhile
		\State {forward lowest FSV flow from DA2GC over SA2GC}
		\EndWhile
		\EndIf
		\EndProcedure

		\Procedure {change in capacity over SA2GC}{new capacity}
		\If {new capacity $>$ old capacity}
		\While {DR of the highest FSV dropped flow $\leq$ RC over SA2GC}
		\State {forward dropped flow with the highest FSV over SA2GC}
		\EndWhile
		\Else
		\While {RC over SA2GC $\leq$ 0}
		
		\State {drop the flow with the lowest FSV from SA2GC}
		
		\EndWhile
		\EndIf
		\EndProcedure
		%
		
	\end{algorithmic}
	\label{alg:procedures}
\end{algorithm}

%
%

\section{Simulation Setup} \label{sec:sim_setup}

The above described traffic and link models were implemented in \ac{ns-3} v3.28 \cite{ns-3}. By default, it offers OpenFlow version 0.8.9. However, we upgraded it to OpenFlow version 1.3 \cite{SDN-ofswitch}, which adds the dpctl utility. It is a management utility to control over the OpenFlow switch, for sending/receiving messages between controller and switch as well as flow meters. The implementation includes the above described traffic and link model as well as a forwarding logic for flows based on one of the outlined forwarding schemes. The logic of the forwarding algorithm is implemented in the \ac{SDN} controller. 

Default queues are limited in size and cannot be modified in OpenFlow 1.3, hence, a queue concept was implemented in the traffic flow sets maintained by the controller. \ac{FIFO} queuing technique forwards packets entering first and hence leaving the queue first. This could imply that packets from high data rate flows, which arrive continuously, such as those from Video, would be able to fill up the queue with the flows having  non-continuous traffic generation. Low data rate flows, such as VoIP, would not be able to turn on and compete for the queue capacity. This could lead to a performance degradation for the VoIP flows. To account for such effects, the sets maintained in the controller are, additionally to the sorting based on a flow's FSV, sorted based on the arrival time of the flows. This means that the \ac{FIFO} principle is applied on the sets on a per flow basis. It is also fitting with the ``continuous" connections assumed for the simulation. Therefore, contrary to what could intuitively be assumed, i.e. that flows with the same FSV achieve the same performance, the actual performance achieved is also conditioned by the arrival time of the flow.

We consider an aircraft with $107$ passengers, which includes $95$ economy, $6$ business, and $6$ first class passengers as defined in Section \ref{sec:trafficmodel} for a one-hour flight. The \ac{QoS} metric is chosen to be the percentage of  dropped packets and delay during a one-hour flight. For dropped packets, \ac{QoS} levels are assumed to be $1$\%, $2$\% and $10$\% in \ac{VoIP}, Video and Web applications, respectively. Mission critical flows are again assumed to have no tolerance for delay, i.e. only tolerating the minimum delay possible – which is in this case only when transmitted over DA2GC. MTC is assumed to tolerate delay. The only application from the PODD domain which does not tolerate the delay levels experienced over SA2GC, is VoIP \cite{cisco2}. To quantize the delay QoS threshold, the amount of time VoIP flows spend on SA2GC shall not exceed $1$\% for all travel classes. For the link capacities, we assume $100$ Mbps  per DA2GC and SA2GC spot as this is the order of magnitude of currently available capacities. We have increasing and decreasing user activity level during the flight.

\section{Simulation Results} \label{sec:results}


\subsection{Dropped Packet Performance}

\begin{figure}[!t]
	\centering
	\includegraphics[width=.45\textwidth]{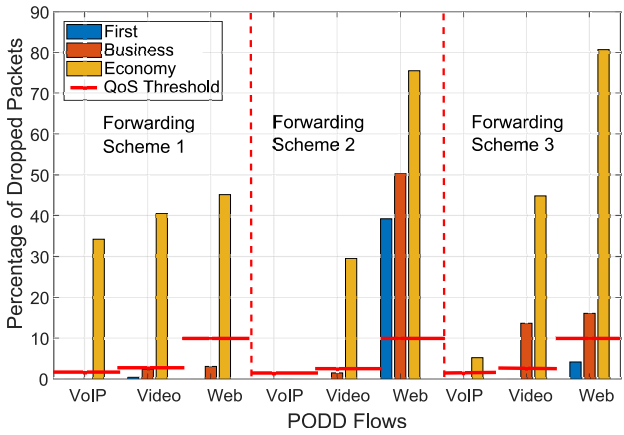}
	\caption{Percentage of dropped packets for traffic flows from the PODD domain for different forwarding schemes.}\label{fig:drop}
\end{figure}

Figure \ref{fig:drop} shows the percentage of dropped packets throughout the flight when different forwarding schemes are applied. For forwarding scheme 1, packet loss is observed in all travel classes, however, it is correspondingly lower in domains belonging to a higher priority class. Hence, first class has the lowest packet loss, followed by business and economy class. QoS thresholds are being exceeded for all flows from economy class, and for Video from business class. Applications with the same priority from the same travel class should have the same packet loss. However, due to the assumed FIFO queuing which is implemented on the traffic flow sets handled by the controller, it is not the same for all applications. 
At some points in the simulation, flows which arrive at the link first, occupy the link and block other flows arriving later. Although some VoIP, Video and Web flows are assigned with the same start times, the ns-3 event loop cannot start them at the same time. There is a slight difference between arrival times of flows with the same start time. Hence, some flows arrive to the switch first – in this case VoIP flows. Therefore, a VoIP flow arriving first will affect the performance of Web flows arriving later due to the ns-3 event loop. Additionally, not all applications have the same number of traffic flows. While VoIP and Web applications have the same number of flows, there is a much larger number of Video flows arriving at the switch. Given the higher number of flows from Video, the time between starting two VoIP or Web flows is higher than the one between starting two Video flows. Multiple Video flows arriving before the next VoIP/Web flows also affect their performance. Flow timeout after drop-out is not considered due to the CBR traffic assumption of flows aggregated behavior of PODD. 

For forwarding scheme 2, flows with a high delay requirement are performing better than those with a lower one. It is able to meet the packet loss QoS requirements of only VoIP flows from all classes at the expense of the performance of Web flows. Nevertheless, a more advantageous distribution of packet loss (i.e., for lower priority classes) among the travel classes is achieved.

For forwarding scheme 3, only first class applications are satisfied. there is an overall improvement in the performance of Web flows  compared to the performance of forwarding scheme 2, along with a performance degradation of VoIP and Video flows. However, this degradation is not as significant as the improvement of the performance of Web flows.
\subsection{SA2GC Link for VoIP Flows}

\ac{VoIP} flows are delay intolerant and their requirement in terms of delay cannot be satisfied if they they are forwarded over \ac{SA2GC} link. Hence, we quantize \ac{QoS} threshold as $1$\% of the time VoIP flows spend on SA2GC link.

\begin{figure}[!t]
	\centering
	\includegraphics[width=.45\textwidth]{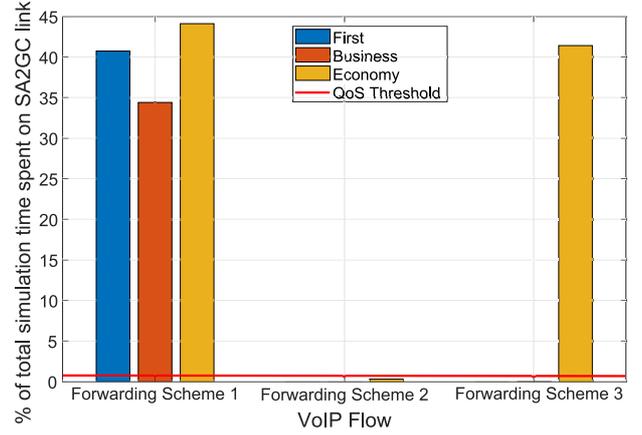}
	\caption{Percentage of total simulation time spent on the SA2GC link for VoIP flows from each travel class for different forwarding schemes.}\label{fig:sa2gc}
\end{figure}

Figure \ref{fig:sa2gc} shows the percentage of total simulation time spent on the SA2GC link by VoIP flows from each travel class. The thresholds are exceeded for all classes if forwarding scheme 1 is used. The performance of the first class is worse than the business class since the performance of the VoIP flow from first class is being affected by a higher number of Video flows with the same FSV. Those occupy the link in overload situations as the VoIP flow must leave the link first. The flows from the business class are based on the forwarding scheme sorting and allocation to links closer to being dropped. 
Overall, the queuing technique used causes some ``counter intuitive" effects. The effects are most vivid when there are many flows with the same forwarding scheme value, which happens with this forwarding scheme.

Forwarding scheme 2 shows that the thresholds are not exceeded for any travel class. It aids a better performance for high delay requirement flows even with their lower priority. 

For forwarding scheme 3, the delay QoS requirement is not met for VoIP flows from the economy class. The performance degradation of VoIP flows propagates also to the times these flows are forwarded over the SA2GC link due to the additional considered parameter. To achieve an improvement of performance of Web flows, the performance of the other flows must be diminished.

\subsection{Cache Hit Ratio}

Concerning the required cache hit rates, we re-run the simulations with each forwarding scheme for different traffic and link conditions. In each run, the seeds of random number generators are updated, which effectively determine the duration of ON/OFF intervals of traffic flows, as well as the type and size of spots traversed. 
For each forwarding scheme, a total of $295$ runs are executed. For a given random number generator seed, multiple simulation runs are performed to find the required cache hit rate for the conditions generated with the seed. 

We calculate the cache hit rate as the percentage of Web and Video traffic which needs to be stored on the aircraft to satisfy \ac{QoS} requirements of the all traffic flows. This cache hit rate is assumed to be required for  the generated aircraft route and input traffic conditions in their respective simulation runs. The simulation time was scaled for running the simulation with multiple input conditions. This leads to scaling the ON/OFF intervals of applications, as well as the intervals at which changes in link conditions occur.

\begin{figure}[!t]
	\centering
	\includegraphics[width=.40\textwidth]{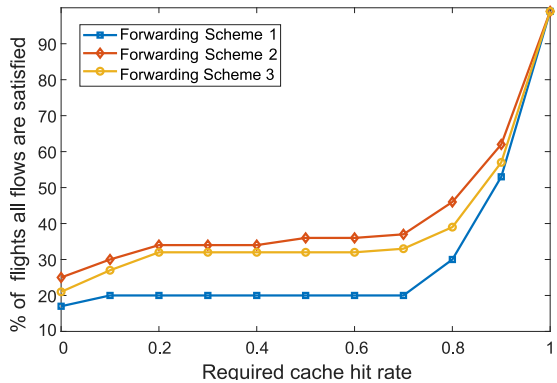}
	\caption{Percentage of simulated flights where  \ac{QoS} requirements of all flows are satisfied with respect to required cache hit rates for forwarding schemes.}\label{fig:hit_rate}
\end{figure}

Figure \ref{fig:hit_rate} shows the cumulative distribution function of the percentage of simulated flights where \ac{QoS} requirements of all flows are satisfied with respect to required cache hit rates for the forwarding schemes. A $100$\% cache hit rate is required to have all flows satisfied for any forwarding scheme. However, this is hardly feasible. The next most probable scenario is $90$\% cache hit rate requirement, where all flows are satisfied in more than $50$\% of flights within the simulation runs. Nevertheless, a small cache hit rate such as $0.2$, can already improve the performance in $20$-$30$\% of the routes as compared to no caching.

Overall, it can be stated that the performance when only considering priority for the forwarding scheme calculation is good only for flows with a correspondingly high priority value. It is also the case that a much higher cache hit rate is required to make up for the low performance of low priority flows. With forwarding schemes which consider additional parameters, such as delay requirement and the number of times a flow was dropped, the performance is improved for flows which have a correspondingly higher value in these additional parameters. This however, happens at the expense of degrading performance of other flows. Furthermore, the required cache hit rates to make up for that performance degradation with forwarding schemes 2 and 3 are much lower than those which are required with forwarding scheme 1. Further research is needed to investigate under which conditions the metrics should be combined to achieve a longer duration during the flight where all flows are satisfied.

\section{Conclusion}\label{sec:conclusion}
In this paper, we propose models to represent the traffic flows within the aircraft with variations of \ac{SA2GC} and \ac{DA2GC} links. We implement these models in our simulation framework using  \ac{ns-3} with OpenFlow. A forwarding logic is adopted with three different forwarding schemes based on a weighted combination of priority, delay requirement and the number of times flow was dropped. From a set of simulation runs, it has been shown that the performance with the forwarding scheme 1 configuration, which relies only on traffic flow priority, is worst in all simulation scenarios. Forwarding scheme 2 and forwarding scheme 3 on the other hand are comparable, and yield better performance, since they are based on more parameters. However, forwarding scheme 2 has better QoS performance in PODD flows than forwarding scheme 3. Overall however, it is also shown that often, regardless of which forwarding scheme is used, QoS thresholds of the flows are exceeded with the current capabilities of \ac{DA2GC} and  \ac{SA2GC}. Therefore, caching is needed as a fallback technology. However, even for the cache hit rate lower than $0.9$, all flows are satisfied in around $50$\% of simulated flights. From the results, $100$ Mbps capacity per BS and satellite is not enough for the modeled aircraft traffic and user activity up to $100\%$. Hence, future directions include emulation of scenarios with higher link capacities and an optimal combination of metrics for the forwarding scheme with a queuing technique inherent to flow characteristics.



\section*{Acknowledgment}
This work is supported in part by EIT Digital ICARO-EU (Seamless Direct Air-to-Ground Communication in Europe) Project. We thank Michal Vondra for his valuable contributions and technical discussions.

\bibliographystyle{IEEEtran}
\bibliography{bibliography2}

\begin{acronym}[CSMA/CA~]
 
 \acro{IFBC}{In-flight broadband connectivity}
 \acro{BS}{base station}
 \acro{ACD}{Aircraft Control Domain}
 \acro{AISD}{Airline Information Service Domain}
 \acro{MTC}{Machine Type Communication}
 \acro{DA2GC}{direct air-to-ground communication}
 \acro{A2AC}{air-to-air communication}
 \acro{EAN}{European Aviation Network}
 \acro{SA2GC}{satellite air-to-ground connection}
 \acro{SDN}{software-defined networking}
 \acro{ns-3}{network simulator 3}
 \acro{QoS}{quality of service}
 \acro{A2G}{air-to-ground}
 \acro{VoIP}{Voice over Internet Protocol}
 \acro{CBR}{constant bit rate}
 \acro{MTC}{Machine-Type Communication}
 \acro{PIESD}{Passenger Information Entertainment Services Domain}
 \acro{PODD}{Passenger Owned Devices Domain}
 \acro{FRD}{Flight Recorder Domain}
 \acro{HMD}{Health Monitor Domain}
 \acro{FIFO}{First In First Out}
\end{acronym}

\end{document}